# Ultrashort Pulse Generation in Modeless Laser Cavity

Dan Cheng, Yujun Feng, Meng Ding, Debasis Pal and Johan Nilsson, *Fellow OSA*

*Abstract*—We demonstrate experimentally that random phase modulation of an erbium-doped fiber ring-laser by an intra-cavity electro-optic phase modulator did not inhibit ultrashort-pulse operation. Stable and self-starting ultrashort-pulse operation with a single pulse circulating in the cavity was achieved even when the phase modulator was driven with random sequences sufficiently fast and strong to render the laser cavity modeless, in the sense that heterodyning of the laser output did not show any spectral lines corresponding to a mode spectrum. No significant change in measured pulse characteristics was observed, compared to conventional mode-locking in the unmodulated cavity. The insensitivity to the random phase modulation is expected, given the lack of phase-sensitive elements in the cavity.

*Index Terms*—Fiber lasers, laser mode locking, optical pulses, phase modulation.

## I. Introduction

ULTRASHORT pulse lasers (USP lasers, USPLs) can operate in a wide range of regimes and laser cavities. Many types of USPLs produce trains of output pulses resulting from the partial outcoupling of a pulse circulating in the cavity. Insofar as the output is periodic, the optical spectrum comprises discrete lines, and the pulses are often described as a superposition of longitudinal cavity modes with a fixed phase relation, i.e., phase-locked or mode-locked, where frequency-pulling renders also the modal frequency-spacings the same [1, 2]. "Mode-locked laser" and "ultrashort-pulse laser" are then sometimes used synonymously for this type of USPLs. On the other hand, a pulse circulating without interruption in a cavity with periodic amplification and outcoupling of energy is an easy-to-understand, intuitive, picture which does not need to rely on the existence of cavity modes with their characteristic round-trip phase of an integer number of 2 $\pi$. The description in terms of phase-locking of modes then seems farfetched and perhaps even inappropriate for USPLs with no clear roundtrip cavity phase. This includes USPLs with an uninterrupted circulating pulse such as those with frequency-shifted feedback [3]. It has also been suggested that also for more conventional USPLs that do exhibit spectral lines, mode-locking is a misnomer [4] and to quote Arissian and Diels, "one can reasonably argue that a femtosecond pulse circulating in a long cavity has no more coupling with the modes, hence the notion of 'mode-locking' should be a misnomer." [2] Thus, although it is clear that mathematically, Fourier transformation of a periodic pulse train leads to spectral lines, their relation to cavity modes is less clear. Alternatively, many USPLs are described exclusively [4] or at least predominantly in the time domain, and the pulse formation is described, e.g., in terms of effects such as a saturable loss together with self-phase modulation and group velocity dispersion [5, 6]. Such a time-domain description does not need to rely on cavity modes with specific roundtrip phases, which suggests that their existence is not essential even for conventional USPLs. Indeed, those pulse-forming effects do not depend on the value of any reference phase of the optical field, nor on the carrier–envelope offset phase. (We do not consider the regime of few-cycle pulses, which may be different. See, e.g., [7].).

In this work, which extends our previous conference paper [8], we present an erbium-doped fiber ring-laser in which a semiconductor saturable absorber mirror (SESAM) and nonlinear polarization evolution cause an ultrashort pulse to be circulating and outcoupled to form a pulse-train. An electro-optic phase modulator (EOPM) is also spliced into the cavity to investigate the effect of random modulation of the cavity roundtrip phase on the USPL. First, when the EOPM was not driven, this operated as a conventional passively mode-locked USPL with distinct signal-reference beat notes in a heterodyned spectrum, corresponding to the optical spectrum as down-converted to the radio-frequency (RF) domain. We then used the EOPM to impose random changes of the phase which were large and fast enough to make the cavity modeless in the sense that the spectral lines (i.e., beat notes) disappeared. Still, pulses were formed. Pulse characteristics measured with optical spectrum analyzers, oscilloscopes, an RF spectrum analyzer, and an autocorrelator were the same or nearly the same with and without random phase modulation even when the phase imposed by the EOPM was nearly uniformly distributed over the range 0 – 2 $\pi$ rad. This confirms that the measured USPL characteristics are largely unrelated to the existence of cavity modes, and if there is any coupling to the

This work was supported by the Air Force Office of Scientific Research under Grant FA9550-14-1-0382. Dan Cheng is supported by Beijing Jiaotong University (16111004). (Corresponding author: Yujun Feng.)

Yujun Feng, Meng Ding, Debasis Pal and Johan Nilsson are with the Optoelectronics Research Centre, University of Southampton, Southampton, SO17 1BJ, U.K. (e-mail: fabius769@163.com; m.ding@soton.ac.uk; d.pal@soton.ac.uk; jn@orc.soton.ac.uk).

Dan Cheng is with the Optoelectronics Research Centre, University of Southampton, Southampton, SO17 1BJ, U.K. and with School of Electronic and Information Engineering, Beijing Jiaotong University, Beijing 100044, China, (e-mail: DanCheng621@gmail.com ).

cavity modes in the unmodulated case then its effect on the measured characteristics is negligible. Both the USPL and a similar free-running continuous-wave (CW) laser showed that the power of heterodyned signal-reference beat notes decreased with increasing amplitude of the random phase sequence, down to a fraction of 20% or less of the total power. We explain the residual low power primarily in terms of measurement noise and limited trace lengths.

## II. ERBIUM-DOPED FIBER LASER WITH PASSIVELY INDUCED ULTRASHORT-PULSE OPERATION AND WITH RANDOMLY DRIVEN INTRA-CAVITY PHASE MODULATOR

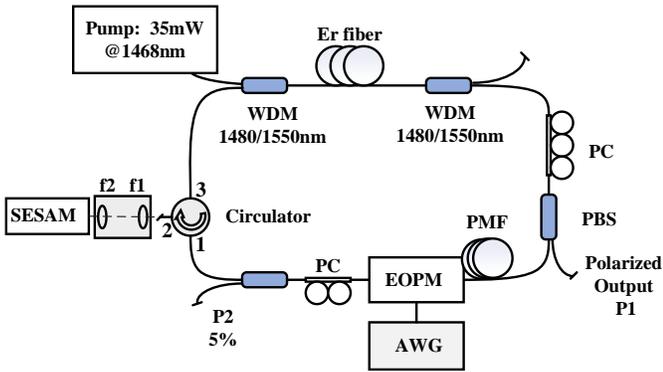

Fig. 1. Schematic diagram of the erbium-doped fiber laser with passively induced USP operation and with an intra-cavity electro-optic phase modulator (EOPM). WDM: wavelength-division multiplexer, EDF: erbium-doped fiber, PBS: polarization beam splitter, PC: polarization controller, AWG: arbitrary waveform generator, P1: port 1, P2: port 2, SESAM: semiconductor saturable absorber mirror, PMF: polarization-maintaining fiber.

The experimental setup of the laser is depicted in Fig. 1. Except for the SESAM arrangement, fiber and fiberized components are used throughout. It is based on the mode-locked laser presented in [9] (i.e., the version without a tunable filter). The SESAM ensures that pulsing self-starts whereas nonlinear polarization evolution adds additional pulse shaping [9]. We modified this laser by adding the EOPM (iXblue, 10 GHz, $V_\pi$ = 4.4 V @ 50 kHz, insertion loss ~3 dB), which increased the loss, length, and anomalous dispersion of the cavity. The other components remained the same. In detail, a 1.8-m-long $Er^{3+}$-doped fiber (EDF) (Fibercore, mode field diameter 5.5 μm, NA 0.22, peak absorption 37 dB/m at ~1530 nm) was pumped by a diode laser centered at 1470 nm with up to 150 mW of power launched through a wavelength division multiplexer (WDM). A second WDM at the other end of the EDF coupled out unabsorbed pump power. A polarizing beam splitter (PBS) introduced polarization-dependent loss (PDL) and polarized linearly the input to the subsequent EOPM. The PBS also coupled out the orthogonal polarization from the cavity through port 1 (P1). The outcoupling corresponds to a cavity loss which depends on the polarization state, which will to some degree self-adjust to reduce the loss. The EOPM was driven by an arbitrary waveform generator (AWG) at 250-MSa/s with ~4-ns risetime (AFG31052, Tektronix). Except for the PBS and EOPM (including the pigtails between them), all components and fibers (predominantly SMF-28) were nominally polarization-independent, without significant PDL. This includes the input pigtail of the PBS and the output pigtail of the EOPM. The EOPM was followed by a tap coupler with 5% outcoupling through port P2 and a three-port circulator. A SESAM with 14% modulation depth, glued to a gold-plated cylinder, was coupled to the circulator's intermediate port (#2) by two lenses. Their focal lengths led to a suitable energy fluence on the SESAM. These were estimated to 17 mm (f1, Thorlabs C260 TME-C) and 12 mm (f2, Thorlabs C220 TME-C) at 1560 nm (chromatic dispersion makes the focal lengths slightly longer than the values specified at shorter wavelengths). Two polarization controllers (PCs) controlled the effect of the nonlinear polarization evolution and the polarization-dependent loss in the PBS. The total cavity length was ~16.2 m and the dispersion approximately –0.26 ps$^2$, estimated from the data in [9] and the added fiber length. The primary diagnostics included an optical spectrum analyzer (OSA, Ando AQ6315E) to measure optical power spectra. Its smallest resolution bandwidth is specified to 50 pm (~6 GHz at 1560 nm). A four-channel 50 GSa/s oscilloscope (Tektronix DSA72004B) with 20 GHz bandwidth (extended through signal processing from the hardware bandwidth of 16 GHz) captured temporal traces from biased InGaAs detectors with bandwidth 15 GHz (EOT ET-3500F), 22 GHz (EOT ET-3600F), and, occasionally, 1.2 GHz (Thorlabs DET01CFC). The total effective bandwidths can be estimated according to $f_{tot}^{-2} = f_{osc}^{-2} + f_{det}^{-2}$ to around 12 GHz and 15 GHz, respectively, for the two fast detectors. Their rule-of-thumb risetimes, $0.35/f_{tot}$, become 29 ps and 23 ps. The captured traces are typically 4 μs long, which leads to a spectral resolution of 250 kHz. We used these instruments and settings unless otherwise stated.

We first ran the USPL without phase modulation, whereby it operates as a conventional passively mode-locked USPL. With the two PCs properly adjusted, we generated stable ultrashort pulses centered at ~1560 nm at 12.7 MHz pulse repetition frequency (PRF) (single-pulse operation with cavity round-trip time 78.7 ns), for a pump power of ~20 mW. All measured characteristics and other observations were as expected for this type of USPL and consistent with the results in [9]. Then, in order to study the impact of the modulation of the roundtrip phase on the pulse formation, the AWG was set to drive the EOPM with a 10-V peak-to-peak random sequence of 0.4 ms duration (100,000 points at 250 MSa/s), periodically repeated at a modulation repetition frequency (MRF) of (0.4 ms)$^{-1}$ = 2.5 kHz. The waveform was generated in an infinite loop, continuously without any gap or other glitch between repetitions. This was verified with a glitch-triggered oscilloscope (Keysight MSOX4154A). Furthermore, 0.4 ms is expected to be long enough to avoid any effects of accidental resonances between the PRF and the MRF. A section of the AWG output is shown in Fig. 2(a) over 80 ns, i.e., just over one roundtrip. This was measured with the 20-GHz oscilloscope and recalculated to the phase $\varphi_{EOPM}$ induced by the EOPM with $V_\pi$ = 4.4 V. The phase varies within a range of ±1.13 π rad (i.e., over a total range of 2.27 π rad = 7.13 rad). Fig. 2(b) shows the amplitude spectrum of the phase modulation, found by Fourier-transforming a 40-μs long section of the voltage trace measured by the oscilloscope. The double-sided bandwidth becomes around 200 MHz (full-width at half-maximum, FWHM). The DC component has been discarded, since it does not contribute to the phase modulation. Fig. 2(c) shows the distribution of changes in $\varphi_{EOPM}(t)$ in one roundtrip time, i.e., $\varphi_{EOPM}(t) - \varphi_{EOPM}(t-80\text{ ns})$, as folded into the range [0, 2π]. This was evaluated every 1 ns for a 4-µs trace. The EOPM phase changes in one roundtrip are nearly uniformly distributed over [0, 2π]. This suggests that the variations are fast and large enough to eliminate any cavity modes as characterized by a roundtrip phase of an integer multiple of 2π rad. The pulse-to-pulse coherence, evaluated from the phase trace as $|<\exp(i\varphi_{EOPM}(t))\exp(-i\varphi_{EOPM}(t-80\text{ ns}))>|$ where the averaging $<\bullet>$ is evaluated as a temporal average, becomes 0.033 for a modulation range of 2.27 π. The calculated coherence reduces to this value already for ~5 ns of delay and remains constant for larger delays, which is consistent with random modulation with 200 MHz bandwidth.

Optical pulse spectra were measured at port P2 with the OSA at 0.05 nm (~6 GHz) resolution, as shown in Fig. 2 (d), with and without phase modulation. We repeated the measurement three times to verify consistency. The characteristics of a ML spectrum with Kelly sidebands are present both with and without phase modulation. There is a noticeable change in the spectrum. This is repeatable but quite small, ~0.1 dB within the 3-dB spectral bandwidth of ~5.45 nm (670 GHz). The deviation remains within ~0.2 dB even 35 dB below the peak, except at the Kelly sidebands, where it reaches ~1 dB. This is primarily caused by slight wavelength changes of the sharp sidebands rather than by a change in their peak values. Fig. 2 (e) shows the corresponding train of pulses in the temporal domain measured with the 1.2-GHz detector at port P1. The random-sequence phase modulation produced no discernible change. Fig. 2 (f) shows single-pulse traces measured with the 15-GHz detector, averaged over ten pulses, with and without random-sequence phase modulation. Again, there is no discernable difference between traces. The peak duration is 40 ps (FWHM). This is comparable to the temporal resolution, so we conclude that the measurement is resolution-limited and that the actual pulses are shorter than that. Since the detection system integrates the signal over the temporal resolution, the peak values of the traces in Fig. 2 (e) and (f) correspond to the pulse energy, which is thus found to be the same with and without random phase modulation. The 670-GHz linewidth corresponds to a transform-limited pulse duration of around 0.5 ps (FWHM), in case of a hyperbolic-secant pulse shape with time–bandwidth product of 0.315. Simulations with RP Fiber Power [10] suggest that the pulses are shorter than 2 ps, and from the Kelly sidebands we estimate a duration of 1.35 ps [11], although the uncertainties in cavity parameters and polarization lead to significant uncertainty in these calculations. Fig. 2 (g) shows the temporal trace of the USPL captured at P1 as the phase modulation turns on and then off after a period of time. There are no obvious transient effects such as loss of USP operation or change in pulse amplitude or number of pulses (e.g., double-pulsing). We also found that ultrashort pulsing self-starts with random phase modulation both on and off. We conclude that the single-pulse state of the laser persisted steadily with and without random phase modulation in the cavity, and did not observe any significant difference in any measured characteristics between the two cases This was true when the phase modulation was switched on and off during USP operation, as well as when the phase modulation was switched on (or not) before the laser was powered on. No difference was observed in self-starting behavior, nor in any other qualitative aspects, which were all as expected.

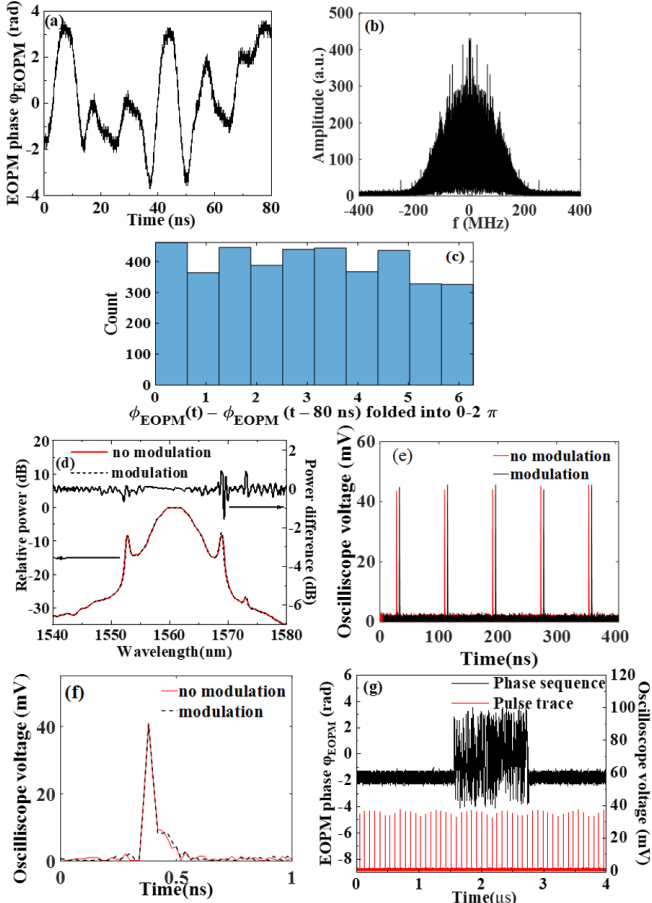

Fig. 2. (a) Time-domain waveform generated from the random sequence and used to drive the EOPM, as recalculated to induced phase difference and shown over 80 ns. (b) Amplitude spectrum (Fourier transform) of a 40-µs section of the waveform shown in (a). (c) Distribution of EOPM phase change in 80 ns (i.e., $\varphi_{EOPM}(t) - \varphi_{EOPM}(t-80\text{ ns})$) as evaluated every 1 ns for a 4 µs trace and folded into the range [0, 2π]. (d) Optical spectrum at unpolarized port P2 for 20 mW pump power with (black curve) and without (red curve) phase modulation. They are nearly identical; the top curve shows the difference. (e) Output temporal traces at polarized port P1 with (black curve) and without (red curve) phase modulation. The pulses are temporally offset for clarity. (f) Zoom-in on a single-pulse trace averaged over ten pulses with and without phase modulation (without offset between them). (g) Pulse traces and phase modulation with random-sequence phase modulation switched on and then off. All curves use either 10 V or 0 V modulation voltage.

### III. MEASUREMENTS ON REINSTATED ULTRASHORT-PULSE LASER

In response to review requests, we reinstated the laser and measured pulse intensity autocorrelations and non-heterodyned RF spectra. We only used the reinstated laser for these measurements. The reinstated cavity was ~1.5 m longer. The PRF was ~11.4 MHz. The linewidth was ~4.5 nm (~550 GHz), centered at 1558 nm and with Kelly sidebands ~15 nm apart, as measured with an Ando AQ6317B OSA at 0.5-nm resolution. The autocorrelator (Femtochrome FR-103XL) had a fiber adaptor on the input port, which was connected to P1 by a 2-m-long PM-980 patchcord. Its dispersion is not known but may be around 10 ps/nm/km, and thus –0.026 $ps^2$ in 2 m. This is negligible for a linewidth of 550 GHz. Fig. 3 (a) shows intensity autocorrelation traces with and without 10-V random phase modulation. The traces (including the small pedestals) are near-identical, and the small difference was not repeatable. The FWHM of the autocorrelation traces is ~1.57 ps. In case of a hyperbolic secant pulse-shape, this translates to a pulse duration of ~1.02 ps and a time-bandwidth product of ~0.56. This is consistent with results reported for an earlier incarnation of this laser [9] and with our simulations. The P1-power was ~0.1 mW, leading to a peak power of ~10 W and a nonlinear phase-shift in the patchcord of a few mrad, which is negligible. Fig. 3 (b) and (c) show signal-signal beat spectra, measured without any heterodyning on P1 with the 15-GHz detector. We used the 125-MHz oscilloscope to capture 125-MSa traces, which were subsequently Fourier-transformed, squared, and folded into a single-sided RF power spectrum. This is also known as the intensity Fourier transform (IFT), as discussed in the next section. The random phase modulation does not induce any change beyond random variations and drifts between measurements. Note that this shows the actual IFT, whereas the figures in other sections show its square. The total RF power was approximately –43 dBm.

By contrast, it was possible to strongly affect the USPL by tuning the frequency of a sinusoidal phase modulation at 10 V peak-to-peak sufficiently close to the PRF. It is plausible that also periodic random modulation with MRF at or near the PRF can perturb USP operation. This may be possible also if the MRF is a sub-harmonic of the PRF. To investigate this, we changed the AWG sample rate and thus the MRF slightly to make the PRF a harmonic of the MRF (estimated to the 4563$^{rd}$ harmonic). There was no observable effect on the optical spectrum or the signal-signal beat spectrum as measures with an RF spectrum analyzer (HP 8560E). This was true also if the MRF was varied around this resonant MRF. We also created a short random waveform with period ~90 ns to enable us to frequency-tune the MRF around the PRF. No meaningful change was observed when the MRF was not resonant with the PRF (e.g., offset 100 kHz in frequency.) When the MRF was tuned close to the PRF, the signal-signal beat spectrum did show small changes, but USP operation persisted. However, although the waveform was created with random samples, the short period was only 42 samples at ~480 MSa/s, and different realizations of the random waveform may have a stronger impact on the USPL.

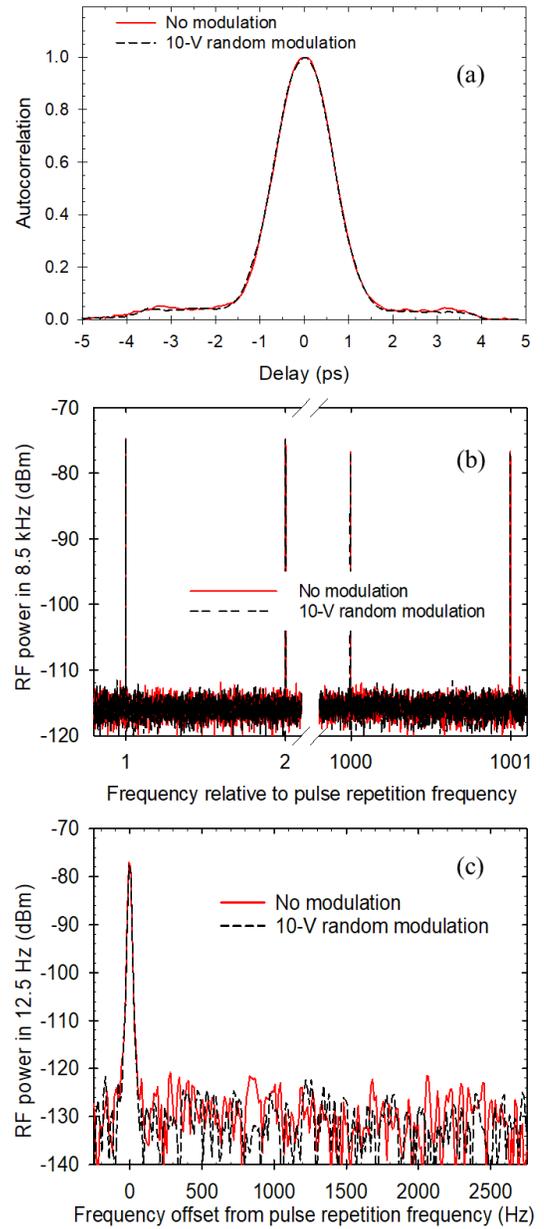

Fig. 3. (a) Autocorrelation traces and sections of signal-signal beat spectra captured over (b) 2.5 ms at 50 GSa/s and (c) 80 ms at 1562.5 MSa/s. In all cases, traces without and with 10-V random modulation are shown.

IV. INVESTIGATION OF THE IMPACT OF INTRA-CAVITY RANDOM PHASE MODULATION ON LASER MODES BY HETERODYNE DETECTION

We used optical heterodyning to reach the MHz-level spectral resolution required to study the modal behavior with intra-cavity random-sequence phase-modulation. If a laser can be described in terms of cavity modes, its rapidly varying field $E(t)$ can be written as

$$E(t) = \sum_k C_k e^{i2\pi v_k t}/2 \qquad (1)$$

where $C_k$ is the amplitude and $v_k$ is the optical frequency of mode $k$. The modes are at both positive and negative frequencies, paired, and enumerated such that $v_k = -v_{-k}$. Ideally, the parameters are time-independent but may vary slowly in time. Furthermore, $E(t)$ can naturally and conveniently be taken to be real-valued. Then, $C_k = C_{-k}^*$, where the star denotes complex conjugate. Because of dispersion, the frequencies of the cavity modes are not necessarily exactly equidistant. However, in case of a conventional single-pulse USPL, $v_k = k \Delta v \pm v_0$, with constant frequency spacing $\Delta v$ equal to the PRF and where $v_0$ is the carrier envelope offset frequency [2], with different sign for positive and negative frequencies. In both these cases, this results in a line spectrum containing all the power at frequencies around ±192 THz in case of emission at ~1560 nm.

Experimentally, the output of the laser-under-test was mixed with that of a stabilized single-frequency (SF) polarized reference laser (IDPhotonics CoBrite-DX4, linewidth < 100 kHz and typically 25 kHz, according to the manufacturer) to down-convert the spectrum to the RF domain. Following square-law photodetection, the signal becomes

$$V(t) = C_{ref}^2/2 + E^2(t) + C_{ref}\sum_k C_k (e^{i2\pi(v_k - v_{ref})t} + e^{i2\pi(v_k + v_{ref})t})/4 \quad (2)$$

Here, the reference field has been expanded as

$$E_{ref}(t) = C_{ref}(e^{i2\pi v_{ref} t} + e^{-i2\pi v_{ref} t})/2 \quad (3)$$

and the laser field as

$$E(t) = \sum_k C_k e^{i2\pi v_k t}/2 \quad (4)$$

in the third term of $V(t)$ (but not in the second term). Furthermore, some terms oscillating with zero mean at twice the optical frequency are neglected to simplify Eq. (2). The signal $V(t)$ is proportional to the instantaneous power, and thus, in the absence of the reference laser, to the instantaneous power of the laser-under-test (proportional to $E^2(t)$ as averaged over the optical cycle). The signal was measured with the oscilloscope and its spectrum $\hat{V}(f)$ obtained as a Fourier transform (FT) [12]. $\hat{V}(f)$ includes the laser's intensity FT (IFT) $\hat{E^2}(f)$, i.e., the FT of $E^2(t)$. This is the spectrum conventionally measured with a RF spectrum analyzer (without mixing with a reference laser), and is also known as the signal–signal beat spectrum. It is centered at zero frequency (DC) with single-sided spectral width equal to the laser linewidth or less. It is independent of the optical phase, and is therefore less interesting for us. Instead, our primary interest is the phase-dependent beating of different spectral components of the signal laser with the reference laser (i.e., the signal–reference beating). Its Fourier transform becomes $C_{ref}[\hat{E}(f - v_{ref}) + \hat{E}(f + v_{ref})]/2$, where $\hat{E}$ is the Fourier transform of the optical wave $E$ and $|\hat{E}|^2$ corresponds to the optical power spectrum. Thus, each of the two terms has a linewidth equal to the optical linewidth of the test laser and provided that $v_{ref}$ is tuned to lie outside the optical spectrum, the two terms will be well separated into a positive-frequency and negative-frequency branch. It is also possible to tune $v_{ref}$ so that the IFT is spectrally separated. Over a spectral range chosen to include only one of the positive-frequency and negative-frequency branches and exclude the IFT, $|\hat{V}(f)|^2$ then becomes the optical power spectrum as down-converted to the RF. Any modes are now resolved, and their power can be evaluated. Note also that the RF spectra we plot are $|\hat{V}(f)|^2$ and may thus include the square of the IFT rather than the IFT itself.

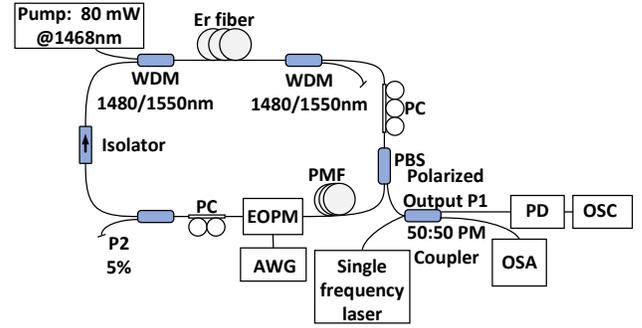

Fig. 4. Experimental setup for free-running continuous-wave laser and heterodyning for investigation of cavity modes.

First, we investigated the modes of a randomly cavity-phase-modulated continuous-wave (CW) laser. Its relatively narrow spectrum makes it simpler to investigate than a USPL, since the power per mode is higher and since it is relatively simple to tune $v_{ref}$ to spectrally separate the IFT from the signal-reference beating and cleanly measure the laser spectrum over its full bandwidth. We also mention that the effect of the Kerr nonlinearity on the roundtrip phase and mode-spacing is negligible in the CW regime. The path-averaged intracavity power may be of the order of 10 mW, and with an average nonlinear coefficient $\gamma$ of ~1 W$^{-1}$ km$^{-1}$, the order-of-magnitude estimate of the roundtrip nonlinear phase becomes 10 µrad. The experimental setup is shown in Fig. 4. Compared to the USPL, the CW laser uses an isolator in place of the SESAM and circulator. Other components remained the same. The cavity is slightly shorter, leading to a mode-spacing of 13.6 MHz. The outputs from the polarized output port 1 (power 2 mW) and the reference laser (power 40 mW) were combined by a 50:50 polarization-maintaining coupler. Fig. 5 shows the resulting spectra $|\hat{V}(f)|^2$ when the EOPM was driven by a random sequence of different amplitudes. For this, 4-µs-long traces (~52.8 roundtrips) were captured by the 15-GHz detector and oscilloscope at 50-GHz sampling rate and then Fourier-transformed. The peak-to-peak amplitudes (i.e., the total voltage spans) were 0 V, 4 V, 6 V, 8 V, and 10 V, corresponding to phase ranges of 0.91 π rad, 1.36 π rad, 1.81 π rad, and 2.27 π rad. The small shift in the signal-reference beat spectrum seen in Fig. 5 (a) - (b) may be caused by small wavelength drifts in the lasers as well as by the modulation. The linewidth is in the range 2 – 4 GHz (RMS) and seems to be systematically ~2 GHz larger with modulation than without. Neither the linewidth increase nor the spectral shift was carefully investigated, so neither can be conclusively attributed to the modulation. Spectra without reference laser, i.e., the pure IFT without any signal-reference beating, are also shown, and labeled "IFT" in Fig. 5 (although the IFT remains present also when the reference laser is on). The IFT extends out to around ±1 GHz, and is narrower than the spectral width of the signal-reference beating. This indicates there are significant variations in the instantaneous phase, which broaden the optical spectrum but leave the IFT unaffected. (In the extreme, the IFT is a single line for pure phase modulation.) For the signal-reference beating, the reference laser was offset by ~7 GHz to keep the beat notes separated from the IFT and still within the measurement bandwidth. The higher reference power further reduces the IFT relative to the signal-reference beat spectrum. The high-resolution spectra of Fig. 5 (c) – (g) show that there is one more or less strong beat note per mode-spacing of 13.6 MHz (cf. Fig. 5 (c)). The beat notes gradually fade for increasing modulation voltages. The spectral sampling of $(4\ \mu s)^{-1} = 250$ kHz means there are 52.8 points per 13.6-MHz mode-spacing. We also note that the cavity dispersion is small over linewidths this small. For example, $-0.26\ ps^2 \times (2\pi \times 10\ GHz)^2 = -1.02$ mrad for 10-GHz linewidth. This is small compared to the $2\pi$ mode spacing and the related resolution of $2\pi / 52.8$ in roundtrip phase (each mode adds $2\pi$ rad to the roundtrip phase). Thus, no deviation from a constant mode-spacing is observed

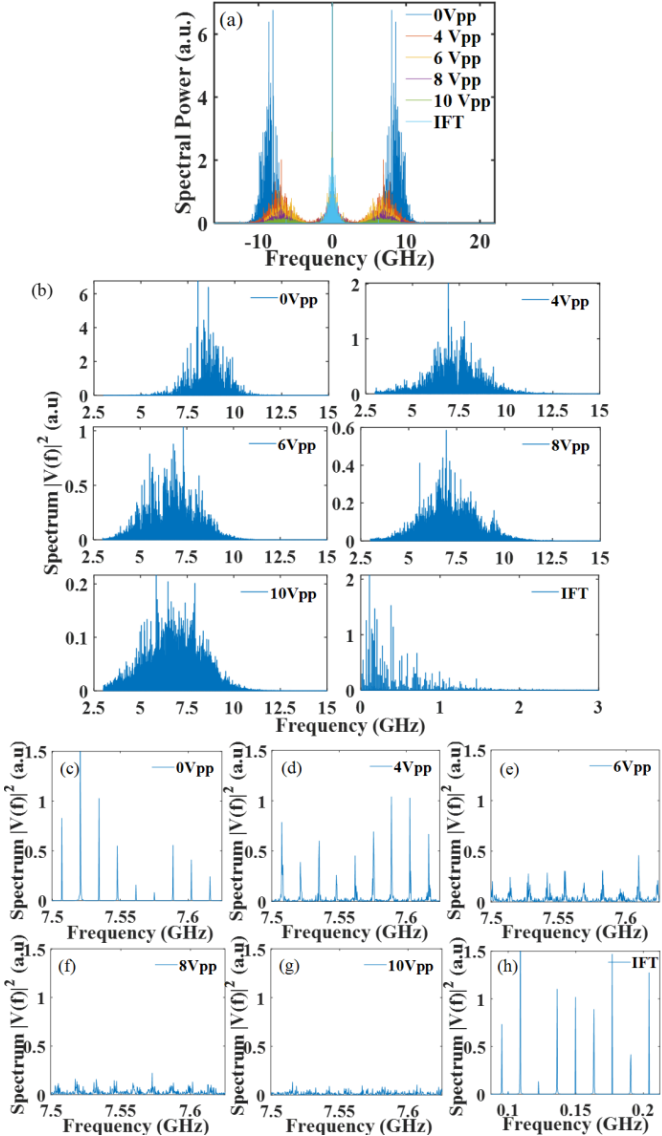

Fig. 5. Spectra $|\hat{V}(f)|^2$ of the CW laser with different levels of random phase modulation for (a) the full bandwidth and (b) zoomed in on the positive branch of the signal-reference beat spectra and on the IFT spectrum. (c)-(g) Further enlargement to a frequency range of about 7.5-7.6 GHz (around the middle of the signal-reference beat spectrum) for increasing random-modulation level as indicated. The amplitude scale is the same in all cases. (h) Enlarged IFT spectrum. All spectra labeled "IFT" were obtained without reference laser and without random phase modulation.

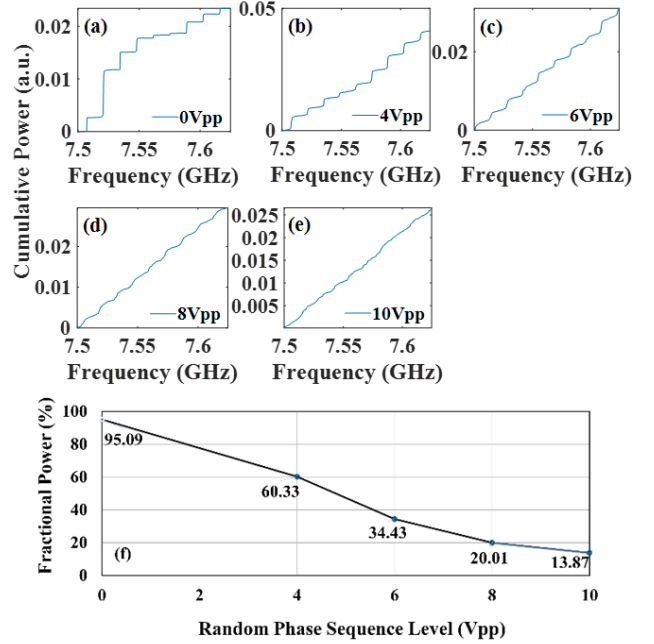

Fig. 6. (a) – (e) Cumulative power relative to total signal-reference beating power of the CW laser as integrated from 7.5 GHz for different phase modulation levels as indicated. The total fractional power in the range shown varies between around 2% and 5% for the different cases. (f) Fractional power as integrated from 5 GHz to 10 GHz in signal-reference beat notes vs. random phase modulation voltage level.

We next evaluated the fraction of power in the beat peaks. First of all, to visualize this fraction, we integrated the spectra $|\hat{V}(f)|^2$ in Fig. 5 in the range 7.5 GHz – 7.62 GHz, where the IFT lines are too small to be observed and the lobe of the signal-reference beating is near its maximum. Fig. 6 (a) – (e) show the spectrally cumulative power over that range. The beat peaks create steps in the curves, and the fractional power is given by the sum of the height of the near-vertical steps relative to the total signal-reference beat power (which includes the power represented by the slant of the more horizontal parts). This allowed for accurate evaluation of the step heights, and thus the fractional power, at low modulation voltages. However, the steps become smaller and less distinct as the phase modulation voltage increases, and increasingly difficult to identify and evaluate. Therefore, to calculate the power in modes even when the background is comparable to the modal peaks, we ensemble-averaged the spectrum into one mode-spacing as follows

$$S_1(f) = \sum_k |\hat{V}(f + k\Delta v)|^2 \qquad (5)$$

The ensemble-averaged spectrum $S_1(f)$ is evaluated for frequencies $f$ varying over one mode spacing $\Delta v$ (from $f_0$ to $f_0 + \Delta v$, where $f_0$ can be chosen arbitrarily). The sum then extends over a chosen frequency range of $|\hat{V}|^2$, namely, 5 GHz to 10 GHz in this case (throughout which the IFT lines were negligible). Random voltage fluctuations in the spectral trace from mode-spacing to mode-spacing sum up to a relatively constant background, whereas any power in modes is periodic in the spectrum and thus sum up to a single spectral peak, which becomes easier to identify. The power in the modes was then evaluated as the power in the highest peak in $S_1$ relative to the total spectrally integrated power. This is conceptually simple but requires that the mode spacing is accurately determined so that the error in $\Delta k \Delta v$ is small compared to $\Delta v$, even when the difference in mode order, $\Delta k$, is close to 400. The mode spacing was determined to about 13.617 MHz. See Appendix 1 for further details.

The resulting fractional power is plotted in Fig. 6 (f). The power dropped from 95.1% to 13.9% when the random modulation level increased from 0 to 10 V (i.e., 2.27 π rad). Imperfections in the distribution (Fig. 2(c)) of the 10-V phase trace (Fig. 2(a)) are evaluated to result in a residual fraction of modal power of less than 4%. Instead, simulations indicated that the 13.9% apparent residual power may primarily be an artefact of the limited trace length and noise.

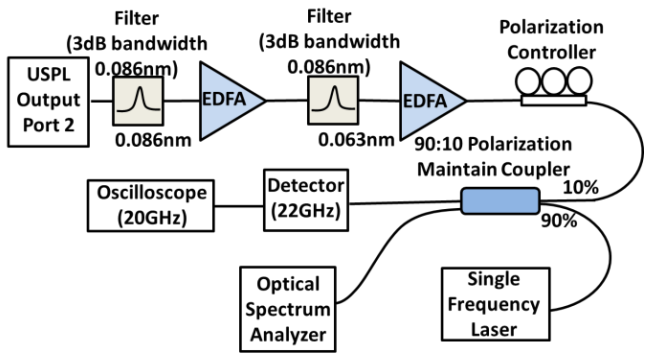

Fig. 7. Experimental setup for heterodyning of the ultrashort-pulse laser.

We next heterodyned the ultrashort-pulse laser using the configuration shown in Fig. 7. For this, we used output port 2. This led to cleaner traces than port 1 did, which was important for the USPL. As a result of re-splicing, the cavity was now slightly longer (PRF 12.3 MHz). The port-2 output power was ~39 μW (pump power ~35 mW). Since the USPL linewidth far exceeds our RF measurement bandwidth, two tunable filters (Alnair-Labs BVF200 and Alnair-Labs BVF200CL), each with 3-dB bandwidth of 0.086 nm, were cascaded to select a 0.064-nm (7.9-GHz) (FWHM) central slice of the spectrum at around 1560 nm. This corresponds to a fraction of ~0.81% of the linewidth of the USPL. The frequency ranges of the IFT and signal-reference beating were restricted accordingly. At the output of each filter, an erbium-doped fiber amplifier (EDFA) boosted the signal power, which became ~0.795 μW (average) at the output of the second filter. Then, a PC adjusted the polarization state of the signal before it was launched into the 10% port of a 90:10 polarization-maintaining coupler to mix it with the reference laser. The resulting beating was measured with the 22-GHz detector. The power of the amplified signal after the combiner was ~2 μW. The power of the reference laser after the combiner was ~4.5 mW to keep it below the power limitation of the photodetector. The transform-limited pulse duration can be estimated to 0.4 / 7.9 GHz = 50 ps. The pulses should thus be slightly longer than our measurement system's response time, though their measured duration may still be affected by it.

Temporal traces of the USPL only (without random phase modulation) as well as the beating with the reference laser without and with 9 V of random phase modulation are shown in Fig. 8 (a). Fig. 8 (b) shows FT spectra $|\hat{V}(f)|^2$ of the temporal traces. The spikes, e.g., at ~12 GHz, are detection artefacts and were removed before further processing of the spectra. The reference laser was offset by ~10 GHz from the center of the filtered spectrum of the USPL. This results in well-separated positive and negative branches of the signal-reference beat spectrum. The linewidth of each branch agrees with the 7.9-GHz bandwidth of the optical filters. Without reference, $|\hat{V}(f)|^2$ has a central lobe (corresponding to the IFT) with width comparable to the width of the beat lobes when the reference is present. Outside the IFT lobe, there is significant noise which even grows to secondary peaks at around ±18 GHz. The IFT spectrum seems relatively well contained within the detection bandwidth, which indicates that we can measure the actual duration of the optically filtered pulses reasonably well (although we plot the square of the IFT, which reduces the apparent spectral width).

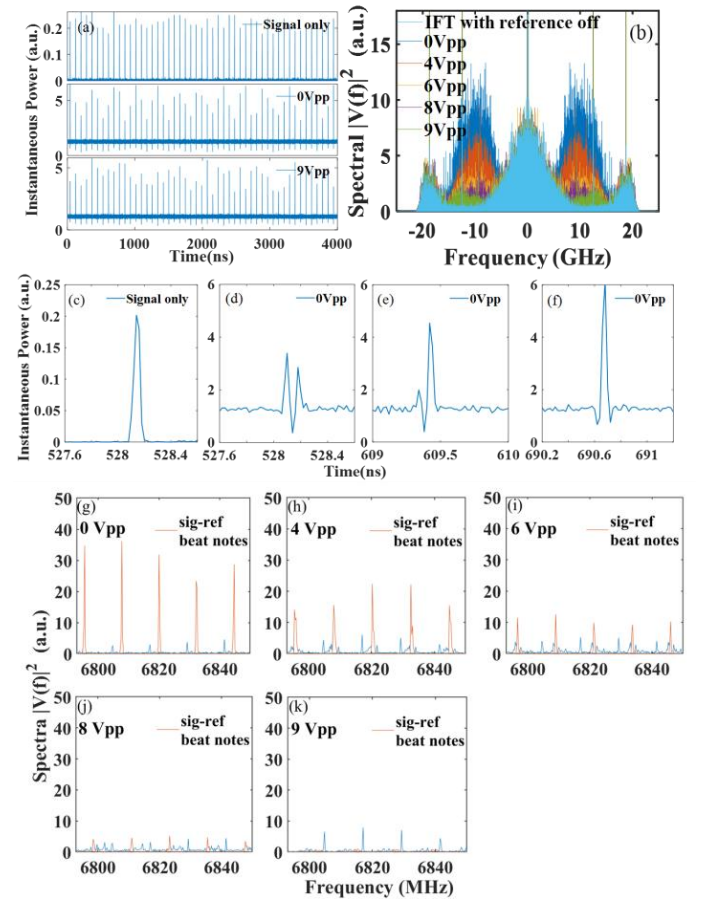

Fig. 8. (a) Temporal traces of optically filtered USPL without heterodyning and without phase modulation (curve labeled "signal only") as well as with heterodyning for random phase modulation voltages of 0 V and 9 V as indicated. (b) Fourier transform $|\hat{V}(f)|^2$ of heterodyned traces for modulation voltages from 0 V to 9 V. The curve labeled "IFT with reference off" is measured without reference laser and without phase modulation. (c) Single-pulse trace without reference laser and without phase modulation. (d)-(f) Zoomed-in examples of a single pulse of the mixed trace with 0 V modulation showing varying interference patterns depending on the relative phase between the USPL and reference laser. (g)-(k) Spectra with frequency scale enlarged to about 6.793-6.855 GHz for modulation voltages from 0 V to 9 V. Each plot contains a single curve, where the parts corresponding to the signal-reference

beat notes are marked in red. The remainder of each curve is blue, and includes the IFT peaks as well as other features outside the signal-reference beat notes. The vertical scale is the same in all plots.

In the heterodyned temporal traces in Fig. 8 (a), the lasers interfere either constructively or destructively, depending on the phase between signal and reference. Without intra-cavity phase modulation, the phase at the pulse peaks drifts at a constant rate relative to the reference phase, leading to a regular pattern of increasing and decreasing peak voltage in Fig. 8 (a). On the other hand, with 9 V of random modulation, the interference at the peaks is constructive or destructive to varying degrees in a random manner. Fig. 8 (c) shows a single pulse without reference laser. The duration is 43 ps (FWHM). This is in good agreement with, and slightly shorter than, the estimated transform-limited duration, which again indicates that the measurement bandwidth is adequate for the filtered pulses. Fig. 8 (d)-(f) show examples of heterodyned single pulses. Each of the heterodyned pulses dips below the baseline level set by the power of the reference laser, indicating destructive interference. Thus, our measurement bandwidth suffices for capturing both constructive and destructive sections of the beating within a single filtered pulse. The positions of the maxima and minima are consistent with a beat frequency of 10 GHz. Note also that here, the phase of the EOPM can be considered constant during the pulse, since the 4-ns risetime of the driver is nearly two orders of magnitude longer than the filtered pulses.

It is also possible to evaluate the peak power from the beating. The highest peaks reach ~5.4 times the reference level. From this, we can roughly estimate the peak power to between 1.2 and 1.6 times the reference power (i.e., between 5.4 and 7.2 mW), where noise and fluctuations in the traces limit the accuracy of the estimate. Given a duty cycle of $43 \text{ ps} \times 12.3 \text{ MHz} = 5.3 \times 10^{-4}$, the average detected signal power becomes 2.9 – 3.8 μW, in fair agreement with the directly measured power of ~2 μW. We note also that perfect cancellation to 0 V should occur occasionally, when the instantaneous pulse power matches the reference power but is in antiphase. This is not seen in our temporal traces. Possible reasons are polarization mismatch, broadband background (e.g., amplified spontaneous emission, ASE), and the limited detection bandwidth. Overall, the characteristics of the temporal trace of the filtered USPL output, without and with mixing with the reference laser, agree with expectations. Fig. 8 (g) – (k) show spectra enlarged to the frequency range of about 6.793-6.855 GHz for intra-cavity random phase modulation voltages between 0 V and 9 V. With increased level of random phase modulation, the amplitude of signal-reference beat notes decreased, as for the CW laser in Fig. 5. Small IFT lines can be seen, too.

We next evaluate the fraction of power that can be assigned to modes from the RF beat spectra. The procedure is similar to that for the CW laser in Fig. 5, although it is more difficult because the noise is higher and the IFT lines extend into the spectral region of the signal-reference beating. First of all, we reduced the noise by selecting a low-noise range of the spectrum, 6.793 GHz to 12.22 GHz, and subtracting the baseline noise (measured without signal) from $|\hat{V}(f)|^2$.

Furthermore, we removed the IFT lines from the spectrum. Although their contribution may be small enough to be neglected (e.g., Fig. 8 (i)), the IFT lines can still interfere with the identification of the modal peaks at high modulation if they are not removed, at least in the lower end of the evaluated spectral range, where the IFT remains relatively large (e.g., Fig. 8 (k)). There are several options for removing them. The signal-reference beat notes and the IFT lines of the USPL have exactly the same frequency spacing ($\Delta v \approx 12.3$ MHz), but are generally offset from each other, since the signal-reference beat frequencies depend on the reference frequency. This can be tuned to ensure the two sets of lines do not overlap. By contrast, the IFT lines lie very precisely on the comb $v_m = m \Delta v$. This is how we identified them and subsequently removed them as described in Appendix 1. As for the CW laser, this requires that we first calculate $\Delta v$ precisely, but this is now straightforward thanks to the distinct pulses of an USPL. Another way to identify the IFT lines is to switch off the reference laser, leaving only the IFT in $|\hat{V}(f)|^2$. Also, if the measurement bandwidth and intermediate frequency (between the signal and reference) are sufficient, one can select a spectral region without IFT lines. Thanks to our large measurement bandwidth, this was possible for us, but limited the spectral range available for the evaluation and was therefore expected to lead to less accurate results. Nevertheless, a detection bandwidth of a few mode-spacings should at least in principle be enough to identify beat notes as well as IFT lines, although noise from the IFT lines (i.e., signal-signal beating) may prove prohibitive at low frequencies.

Following removal of the IFT lines, we evaluated the fractional power in the signal-reference beat peaks for different modulation voltages. Fig. 9 (a) – (e) shows spectrally integrated (cumulative) power for random phase-modulation voltages in the range 0 V – 9 V. This is similar to Fig. 6 (a) – (e) for the CW case. However, to make it easier to visually identify the positions in the signal-reference beat notes and improve the definition of the steps even when the background is comparable to the modal peaks, we ensemble-averaged the spectrum in sections of five mode spacings as follows

$$S_5(f) = \sum_k |\hat{V}(f - 5k\Delta v)|^2 \qquad (6)$$

The sum is from 6.793 GHz to 12.22 GHz with $f$ varying by five mode spacings from 6.793 GHz to 6.855 GHz (as for $S_1$, the origin of $f$ is arbitrary in the sense that any offset is compensated for by the summation index $k$). Fig. 9 (a) – (e) thus plots the spectral integral of $S_5$. The mapping into five mode segments instead of a single mode segment may lead to meaningful differences between the five segments. Other segmentations and frequency spans were investigated, too. Although not discussed here, this proved useful for the development of algorithms. Fig. 9 (f) shows that the fraction of power in the beat notes dropped from ~83.3% to ~20.6% as the random modulation increased from 0 to 2.05 $\pi$ rad (9 V). Note that the summation of $|\hat{V}(f)|^2$ into $S_5$ and $S_1$ does not change the relative power in the signal-reference beat notes.

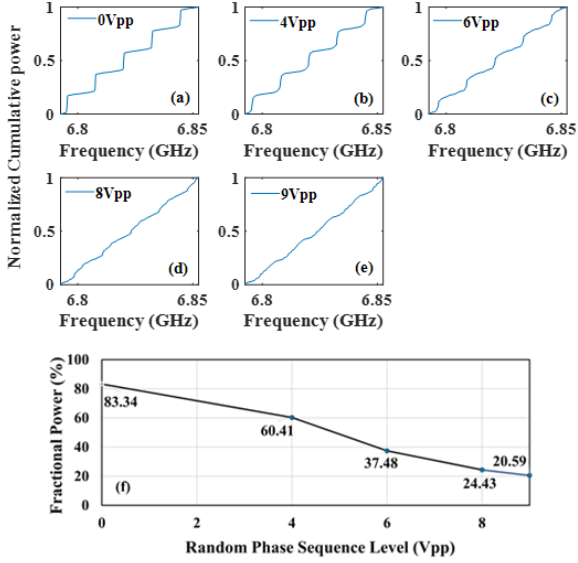

Fig. 9. (a) – (e) Normalized cumulative power of the USPL as evaluated from the range 6.793 GHz to 12.22 GHz and mapped to $S_5$ for different levels of random-sequence phase modulation voltages as indicated. (f) Fractional power as integrated from 6.793 GHz to 12.22 GHz in signal-reference beat notes vs. random phase modulation voltage level.

According to Fig. 9, the fraction of power in the USPL's beat notes is significantly less than 100% even in the absence of phase modulation. Simulations show this may be at least partly explained by the level of noise present in the temporal trace. We simulated the heterodyne detection process for a pulse train without random phase modulation, with and without voltage noise, and then processed the resulting trace as we did the experimental traces. The voltage noise had a normal distribution with zero mean and without correlation from sample to sample (i.e., it was "white"). The average power of the USPL and the RMS level of the voltage noise were set to, respectively, 37% and 1.07% of the reference-laser level, which are similar to the experimental values. Fig. 10 shows the simulated heterodyne temporal trace and spectra $\left|\hat{V}(f)\right|^2$ without and with added voltage noise. The fraction of power in modes was evaluated as for the experimental traces and became 99.5% without voltage noise and 88.4% with voltage noise. We remind the reader that both of these cases are without random phase modulation.

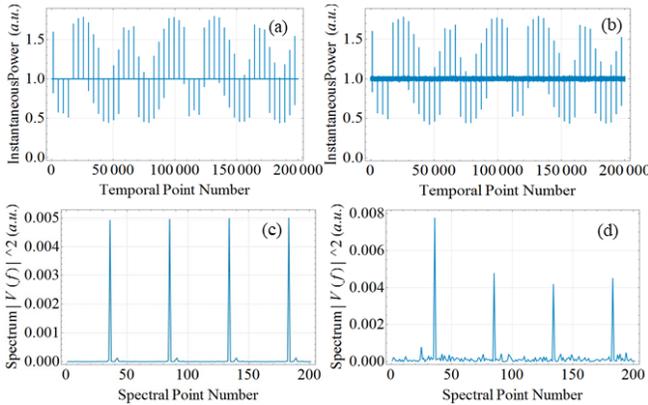

Fig. 10. Simulated heterodyned temporal trace without phase modulation without (a) and with (b) added voltage noise. (c), (d) Corresponding Fourier-transformed spectra $\left|\hat{V}(f)\right|^2$ shown over four mode spacings.

Simulations also show that the apparent residual power in modes at high phase modulations may at least partly be explained by the limited number of pulses in the trace. Fig. 11 shows sections of simulated heterodyne spectra approximately four mode-spacings wide with 0 rad ((a), (c)) and 2 π rad ((b), (d)) random phase modulation for trace-lengths of 49 pulses ((a), (b)) and 49,000 pulses ((c), (d)). At 2 π modulation, the power fraction in modes became 0.18 for the 49-pulse realization, relative to the 0-rad case. This dropped to $1.50 \times 10^{-4}$ for the 49,000-pulse realization. Thus, the power that appears to remain in modes even at high random-phase modulation may be a mathematical artefact related to the statistics of short traces. The imperfect phase distribution with a 9-V modulation range (shown in Fig. 2(c) in case of 10-V modulation range) may also make a small contribution to the residual modal power. In the absence of other imperfections, the contribution was evaluated to 2%. For the pulse-to-pulse coherence with the 9-V trace, this becomes $\left|<\exp(i\varphi_{EOPM}(t))\exp(-i\varphi_{EOPM}(t-80 \text{ ns}))>\right| = 0.081$.

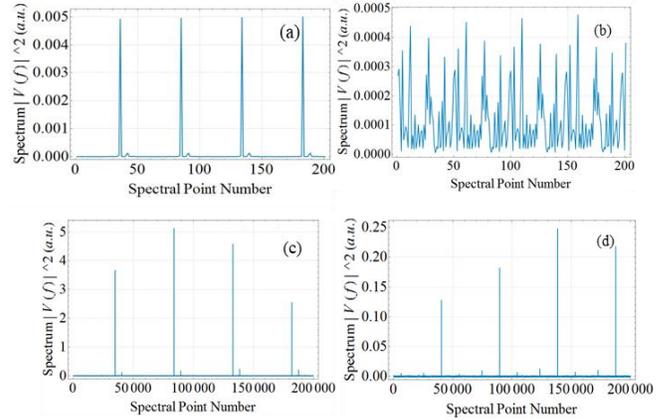

Fig. 11. Simulated heterodyne spectra shown over four mode spacings for 49-pulse traces at the random phase modulation level of (a) 0 rad and (b) 2 π rad and for 49,000-pulse traces at the random phase modulation level of (c) 0 rad and (d) 2 π rad. Note that the highest peaks correspond to signal-reference beat lines in (c), but to IFT lines in (d).

## V. DISCUSSION

We next discuss some additional points which arose during the many reviews of different versions of this paper. Although the oscilloscope captured a trace in as little as 4 μs, the trace was transferred to a computer for off-line evaluation of the mode power (detailed in Appendix 1). The off-line processing allowed us to subtract and average noise, but precluded real-time investigation, and minimization, of the power in spectral lines. However, it may be possible to simplify and expedite the evaluation. We note, for example, that a RF spectrum analyzer was used with heterodyning to characterize a laser with frequency-shifted feedback [13]. At the 12.3-MHz PRF of our laser, 12,300 pulses can be captured in 1 ms, so a RF spectrum analyzer can in principle scan over a few mode-spacings in a low-noise part of the spectrum, ideally where the IFT lines are weak, with acceptable resolution and refresh rate. This will then display signal-reference beat lines, provided the noise is sufficiently low and the lasers are sufficiently stable. In addition, many oscilloscopes, including ours, can also calculate spectra without download, and perhaps fast enough to be considered real-time. The refresh rate of spectra is likely to be limited by the time required for Fourier-transforming a long trace, though we note that the processing speed can be increased by reducing the detection bandwidth and sampling rate, and only calculating the spectrum for a limited range (which in principle can be a band-range over which the noise is low). The oscilloscope approach may rely more on processing with built-in software, which can also limit what is possible real-time *vs.* off-line. Whether an oscilloscope or an RF spectrum analyzer works best is likely to depend on stability and noise, and we note that longer acquisition time allows for lower noise, within limits set by stability and required refresh-rate. With the equipment and approach we used, the stability may well have been adequate for longer acquisition times, and thus possibly for cleaner data, at least following off-line processing. We also expect that the modulation voltage that minimizes the pulse-to-pulse coherence will simultaneously minimize the power in spectral lines. It would then be possible to set up an interferometer and minimize the visibility of fringes (spatial or temporal) and thus the coherence. This can be done real-time, but requires that the interferometer matches the length of the laser cavity to within the transform-limited pulse length (~0.2 mm of fiber for a 1-ps pulse although our optical filtering would increase this to ~10 mm). See, e.g., [14], [15] for further discussions.

The small or negligible difference between the measured quantities of the unmodulated and randomly modulated USPL, without cavity-modes, suggests that the cavity modes of the unmodulated USPL do not affect those quantities. At the same time, mode-based equations that describe USPLs without random phase modulation have, over the years, been derived and verified for a wide range of configurations and operating regimes. We are not aware of anything in our data that contradicts their validity in such cases. On the contrary, even if they were derived without random phase-modulation and with the use of cavity modes, we rather expect that the equations largely remain valid also for our randomly phase-modulated USPL, regardless of if they can still be derived (a solution ansatz for the pulse in the local time frame which compensates for the small modulation of the cavity may be helpful for showing that the equations remain valid). The lack of measured significant changes induced by random phase modulation of our laser, e.g., in the spectrum and average power, as well as in the position of the Kelly sidebands [11], suggests that the validity of at least some equations can be extended to randomly phase-modulated laser cavities.

Some characteristics will change, and presumably mostly for the worse. This includes jitter, where a (random) phase change of $2\pi$ may change the cavity roundtrip time by one optical cycle (5.2 fs), if the group delay due to the modulation of the EOPM is approximately the same as the phase delay. This adds to the jitter of other origin, which may well be much larger than 5.2 fs in our laser as well as in other lasers, if there are no special measures to reduce it. However, jitter at high frequencies (e.g., > 1 MHz) is normally small. This includes pulse-to-pulse jitter, which seems likely to be dominated by the random phase-modulation.

There is also frequency jitter, which through dispersion also adds temporal jitter. The maximum change of frequency induced from one pulse to the next by phase modulation of one wave in 4 ns becomes 250 MHz. We also evaluated the root-mean-square value of the frequency-change to 120 MHz and the average of the absolute value of the change to 90 MHz from a temporal trace such as in Fig. 2 (a) but with high-frequency noise removed. The maximum pulse-to-pulse temporal jitter induced by the change in frequency becomes $\left|-0.26\,\text{ps}^2 \times 2\pi \times 250\,\text{MHz}\right| = 0.41\,\text{fs}$. The overall frequency jitter and associated temporal jitter depends on the ability of the laser to counteract the buildup of the randomly induced frequency changes in the circulating pulse. As suggested by one reviewer, the elastic tape model [16-17] may be able to quantify different properties of a randomly phase-modulated USPL. Further investigations are required to test the extent to which equations describing USPLs remain valid in the presence of random phase modulation.

Our random modulation was sufficient to suppress cavity modes, but USP operation persevered. Still, USP operation can be quite sensitive to perturbations and is expected to end with sufficiently strong random modulation. As it comes to the phase itself, this can always be folded into a $2\pi$ range, and since we already reached this range, stronger modulation is not expected to inhibit USP operation purely due to a larger pulse-to-pulse phase change. As it comes to the random modulation of the group delay and thus the roundtrip time, this will in turn modulate the deposited pump energy and gain recovery in the Er-doped fiber. However, the pump energy modulation will be negligible compared to typical pump ranges over which single-pulse USP operation persists. Still, the modulation can perhaps drive some linear or nonlinear resonance (e.g., in the signal pulse energy) and exceed some nonlinear instability threshold and thus perturb USP operation at a level of modulation that is difficult to predict. Experiments over wider parameter ranges are needed to investigate this, as well as the effect of a stronger random modulation of the frequency.

As it comes to lasers with modeless cavities, different types have been reported in the literature. Of these, lasers with a modulated cavity seem most relevant to ours, and include CW and USP lasers with frequency-shifted feedback induced by an acousto-optic modulator (AOM) (e.g., [13]). Those typically use a constant modulation (i.e., shift) of the frequency, and it was suggested that a frequency-shift of 1% or more of the roundtrip frequency would make the spectrum continuous rather than discrete [18]. In our laser, this becomes only 130 kHz, which is around three orders smaller than the actual frequency shift we induce. In another laser without AOM, the cavity length was modulated [19]. It was estimated that a length modulation of 5% of the wavelength from roundtrip to roundtrip led to modeless operation. This is much smaller than our modulation of around one wavelength, if our $2\pi$ rad modulation is directly converted to length modulation. Furthermore, although their 15.5-kHz sinusoidal modulation had a large amplitude, it was also quite slow with modulation bandwidth four orders of magnitude smaller than ours. The effect of cavity-length changes in the form of mechanical perturbations on spectral lines and mode frequencies is also discussed in ref. [2]. Also fast perturbations are considered, although the brief discussion is not quantitative and the applicability to our laser is unclear. The authors also note that with fast perturbations, the definition of mode frequency is not as simple, which is consistent with our plots of gradually disappearing spectral lines in Fig. 4 and 7.

Noise-like USPLs [14, 20, 21] also lack pulse-to-pulse coherence and thus mode structure. This is a result of nonlinearities which, in contrast to our USPL, leads to an optical spectrum that deviates from that in the conventional mode-locked regime [14, 21]. Although not verified by measurements, other expected differences from our USPL include the pulse-to-pulse spectral variations observed in noise-like lasers [14] and expected non-monotonic changes in the instantaneous frequency, so spectral filtering of a noise-like pulse may result in multiple sub-pulses.

These examples of modeless laser operation are quite different from ours, and without quantification of the fraction of power in modes. Therefore, a comparison to our mode suppression results (e.g., Fig. 5 & 8) is difficult and may not be meaningful. However, we expect that random phase modulation will suppress spectral lines in the same or similar way if the phase-modulation is instead outside the cavity, which may make for more relevant comparisons. External phase modulation, including random phase modulation with noise, is often used to suppress the carrier of continuous-wave signals [22-24], e.g., for suppression of stimulated Brillouin scattering [25]. Generally, the suppression increases with modulation amplitude, though not necessarily monotonically. Modulation with noise nearly completely suppressed the carrier for a phase modulation amplitude of 0.72 $\pi$ (measured as a standard deviation) [26]. Since the modulation was aperiodic and random, the resulting spectrum is continuous. Pseudo-random modulation waveforms (typically maximum-length sequences) can also lead to near-complete carrier suppression for binary phase modulation between 0 and $\pi$ rad (standard deviation 0.5 $\pi$, range $\pi$ rad), although the periodic nature means that weak spectral lines remain. See, e.g., [27, 28]. Engineered modulation waveforms can also reduce the carrier, and was used for the creation of rectangular optical spectra [29]. The range was slightly larger than 2 $\pi$, although the authors (Harish & Nilsson) suggested that a range of ~$\pi$ rad may suffice (as for the pseudo-random case). With 2.27 $\pi$ modulation range, the waveform we use in this paper has a standard deviation of 0.50 $\pi$ rad. Thus, these externally modulated results on carrier suppression are consistent with our suppression of spectral lines through intra-cavity modulation. However, it is worth noting that with intra-cavity modulation, the random phase of the EOPM is imposed on the difference between two successive pulses, whereas with external modulation, it is imposed on each individual pulse. Thus, with external but not with intra-cavity modulation, the pulse phase relative to the unmodulated case is identical to the EOPM phase. (We note also that with intra-cavity random phase modulation, the phase of the successive pulses builds up in a random-walk manner. This needed to be compensated for in some of our simulations. Otherwise, the suppression of modal power and coherence can be over-estimated.) Having said that, the pulse-phase trace with external modulation is perfectly replicated by intra-cavity modulation if the following relation holds:

$$\varphi_{EOPM}^{intra\text{-}cavity}(t) = \varphi_{EOPM}^{external}(t) - \varphi_{EOPM}^{external}(t - 80\ ns) \qquad (7)$$

Thus, the build-up of the pulse-phase can be avoided, but we did not attempt to do that.

Although beyond our primary focus on a laser that lacks spectral lines but still produces a train of pulses, we also mention that a pulse train can exhibit spectral lines even if the pulse train does not originate from a laser cavity, e.g., if it is generated through periodic pulse-carving of a monochromatic wave. Thus, although the absence of spectral lines shows that there are no cavity modes in our laser, there is in general no implication or equivalence between the two. We also expect that an intra-cavity phase modulation that fulfills Eq. (7) would be undone by a matching extra-cavity modulation in the sense that this would lead to a spectral line-structure, even when the cavity-modes have been suppressed. As yet another example, a train of short pulses can in principle be generated by combining independently generated monochromatic waves with uniform frequency-spacing and controlled phase. One may then ask if the spectral lines of the more typical case of a pulse train generated by a conventional USPL with a cavity that does have modes, necessarily correspond to, or couple to, those modes, or if the spectral lines are merely a consequence of the properties of the Fourier transform of a train of phase-coherent pulses. We noted previously that an assumption of a uniform frequency-spacing between cavity modes leads to a 1.02-mrad deviation from a comb of multiples of 2 $\pi$ roundtrip phase, as calculated from the cavity dispersion of –0.26 $ps^2$ for a 10-GHz linewidth (comparable to the bandwidth of our optical filtering). This was not significant for our 4-$\mu$s traces. However, for the full 0.67-THz linewidth, uncompensated dispersion would mean that the pulse duration increases by $\left|-0.26\ ps^2 \times 2\ \pi \times 0.67\ \text{THz}\right| = 1.1$ ps per roundtrip (this is an asymptotic approximation for chirped pulses). It is thus clear that the spectral lines of the comb of our conventionally mode-locked un-modulated USPL deviates from the cavity modes (in the linear regime) so much that the latter cannot be used to describe the pulse train. This disparity is then reconciled by frequency-pulling of modes, which is an essential but arguably somewhat after-the-fact concept if cavity modes are to be used to describe the pulse trains. This underlines that a modal description seems farfetched compared to a time-domain description which does not rely on modes. We would also argue that whereas a frequency-domain treatment is equivalent to a time-domain treatment (insofar as the governing equations can be Fourier-transformed, and regardless of convenience), a treatment in terms of cavity modes may well differ from both.

## VI. CONCLUSION

In conclusion, we have demonstrated a laser with an ultrashort-pulse circulating in a cavity in which phase modulation of the cavity with a random sequence precluded the existence of conventional cavity modes with roundtrip phase a multiple of 2 $\pi$. Experimentally, we used an erbium-doped fiber ring-laser with an intra-cavity electro-optic phase modulator and optical heterodyne detection to verify the suppression of signal-reference beat notes associated with such modes. No significant change of basic laser characteristics such as pulse energy and optical spectrum was observed when the phase modulation was introduced. The results are expected since there is no element in the cavity that is sensitive to the absolute phase or pulse-to-pulse phase variations, and confirm that traditional phase-locked cavity modes are not required for an ultrashort-pulse laser of this type. The cavity modes were also suppressed in a CW laser

APPENDIX

We here describe the procedure we used to identify and remove IFT lines and evaluate the fractional power in the signal-reference beat notes. The starting point is 4-μs-long temporal traces captured by the oscilloscope (200,000 points with 20-ps spacing). The traces were Fourier-transformed to yield $|\hat{V}(f)|^2$ with 250-kHz resolution. The number of spectral points per mode becomes ~12.3 MHz / 250 kHz = 49.2 for the USPL with 12.3-MHz mode-spacing, and 52.8 for the CW laser with 13.6-MHz mode spacing.

1. Determination of mode-spacing $\Delta v$ with high accuracy, so that $v_m = m \Delta v$ is accurate to within a small fraction of the mode-spacing for the full range of modes and frequencies used in the evaluation. The mode spacing was around 12 – 13 MHz and the frequency range considered was around 5 GHz (e.g., 6.793-12.22 GHz, span 5.427 GHz for the USPL)

For the USPL, the mode-spacing (or line-spacing) is equal to the PRF. We determined this from the position of the first and last pulse of each analyzed trace. These are separated by nearly 200,000 points. The pulse trace will be distorted by the beating with the reference laser (see Fig. 9 (c) – (e)) and we estimate the error in the peak position to be comparable to the pulse duration, or around ±40 ps for each peak. Thus, the relative error limit becomes $2 \times 10^{-5}$, so 109 kHz over 5.427 GHz. This is within 0.088% of the mode spacing and smaller than the spectral resolution.

The periodicity is more difficult to determine for the CW laser, since the variations are much less distinct. Therefore, we calculated the autocorrelation of the voltage trace $V(t)$, $C(\tau) = \int V(t)W(t)V(t+\tau)W(t+\tau)dt$, where $W(t)$ is a Hamming window function. This produced distinct peaks in a comb with spacing equal to the roundtrip frequency. The peaks faded gradually for larger $\tau$, but could be identified up to more than 2 μs for a 4-μs trace. From those, the periodicity is easily evaluated. Whereas dispersion is expected to cause the mode spacing to vary in the CW laser, significant dispersion would also cause the autocorrelation peaks to become less distinct, and this was not observed. This further supports that dispersion is not significant, and that the mode-spacing can be evaluated as the inverse of the periodicity also for the CW laser.

The window function eliminated a second, interleaved, comb, which made it more difficult to determine the peak spacing. Since a standard Hamming window produced adequate results, we did not evaluate alternative windows, which may have been able to preserve the amplitude of the first comb to larger values of $\tau$ whilst still suppressing the second comb.

The precise determination of the periodicity is important and non-trivial. Although autocorrelation is well-established and proved adequate, there are several more sophisticated and well-established algorithms for precise determination of periodicity with traces that can be more challenging than ours, and which may be superior for our traces. See, e.g., [30] for a review directed towards astronomy.

We note that the time-base accuracy of the oscilloscope is specified to ±1.5 ppm when new. This error can increase with age, but by less than 1 ppm/year. However, regardless of size, the systematic error this creates in absolute frequency is irrelevant for our analysis. Short-term oscilloscope jitter could be an issue, but is of the order of 1 ps, i.e., much smaller than the 20-ps sample period.

2. A trace was measured without signal and Fourier-transformed to obtain a noise spectrum $|\hat{V}_{noise}(f)|^2$. The noise spectrum was averaged over 100 MHz and then subtracted from the spectra $|\hat{V}(f)|^2$ measured with the signal.

3. The spectra were inspected to identify a region (e.g., 6.793-12.22 GHz) which included most of the signal-reference beat spectrum while exhibiting low IFT and noise.

4. The spectrum was remapped into a single mode-spacing $S_1(f)$ and summed as in Eq. (1). This reduces the noise and facilitates visual inspection of the spectral characteristics of a large number of modes and visual and numerical identification of IFT and signal-reference beat peaks. For this, $|\hat{V}(f)|^2$ is segmented and the segments are summed. The offsets $k\Delta v$ (i.e., the segment boundaries) are rounded to the closest sample point so $|\hat{V}(f)|^2$ does not have to be resampled. Since the mode-spacing is generally not an integer number of points (e.g., ~49.2 points), the segment lengths vary (e.g., between 49 and 50 points). The extra point in the longer traces is discarded.

5. The position of the IFT line within $S_1(f)$ is determined and the line removed. (In $|\hat{V}(f)|^2$, the lines are at $k\Delta v$.)

6. Excluding the IFT line, the highest peak is identified in $S_1(f)$. This is assumed to correspond to the signal-reference beat peak. We checked that it is separated from the IFT line by at least five samples. If not, it may be necessary to select or measure another trace.

7. The average background level in $S_1(f)$ is calculated, excluding the five points closest to each of the IFT and signal-reference beat peak.

8. The IFT is removed by replacing the five samples closest to the IFT line with the average value.

9. The power in the signal-reference beat peak is evaluated as the power in the five samples above the average background level.

10. The total power is evaluated as the sum of the samples in $S_1(f)$ plus the background level in the discarded point (e.g., 49.2 – 49 = 0.2 times the background level).

11. The fractional power is evaluated as the power in the signal-reference beat peak relative to the total power.

ACKNOWLEDGMENT

The randomly phase-modulated laser was modified from a laser originally realized by Dr Luis Vazquez-Zuniga. We thank Dr. Jonathan Price and Dr. Rüdiger Paschotta for helpful discussions and Prof. Radan Slavik for meaningful suggestions. The autocorrelator was provided by Prof. Jayanta Sahu.